\documentclass{article}
\usepackage{arxiv}
\usepackage{geometry}
\usepackage[table,xcdraw]{xcolor}
\usepackage{amsmath}
\usepackage{algpseudocode}
\usepackage{algorithm} 
\usepackage{graphicx}
\usepackage{hyperref}

\algnewcommand\algorithmicforeach{\textbf{for each}}
\algdef{S}[FOR]{ForEach}[1]{\algorithmicforeach\ #1\ \algorithmicdo}

\title{IoT Data Trust Evaluation via Machine Learning}

\author{Timothy~Tadj, Reza~Arablouei, Volkan~Dedeo\u{g}lu\vspace{4pt}\\Data61, CSIRO, Pullenvale QLD 4069, Australia}

\begin{document}
\maketitle

\begin{abstract}

Data trust in IoT is crucial for safeguarding privacy, security, reliable decision-making, user acceptance, and complying with regulations. Various approaches based on supervised or unsupervised machine learning (ML) have recently been proposed for evaluating IoT data trust. However, assessing their real-world efficacy is hard mainly due to the lack of related publicly-available datasets that can be used for benchmarking. Since obtaining such datasets is challenging, we propose a data synthesis method, called random walk infilling (RWI), to augment IoT time-series datasets by synthesizing untrustworthy data from existing trustworthy data. Thus, RWI enables us to create labeled datasets that can be used to develop and validate ML models for IoT data trust evaluation. We also extract new features from IoT time-series sensor data that effectively capture its auto-correlation as well as its cross-correlation with the data of the neighboring (peer) sensors. These features can be used to learn ML models for recognizing the trustworthiness of IoT sensor data. Equipped with our synthesized ground-truth-labeled datasets and informative correlation-based feature, we conduct extensive experiments to critically examine various approaches to evaluating IoT data trust via ML. The results reveal that commonly used ML-based approaches to IoT data trust evaluation, which rely on unsupervised cluster analysis to assign trust labels to unlabeled data, perform poorly. This poor performance can be attributed to the underlying unsubstantiated assumption that clustering provides reliable labels for data trust, a premise that is found to be untenable. The results also show that the ML models learned from datasets augmented via RWI while using the proposed features generalize well to unseen data and outperform existing related approaches. Moreover, we observe that a semi-supervised ML approach that requires only about 10\% of the data labeled offers competitive performance while being practically more appealing compared to the fully-supervised approaches. The related Python code and data are available online\footnote{\url{https://github.com/Tim-Tadj/IoT_datatrust_RWI}}.

\end{abstract}


\section{Introduction} \label{sec:introduction}

The growing industrial use of the Internet of things (IoT) has motivated increased research and development in this field~\cite{DACHYAR2019e02264}. One of the main concerns around the adoption of IoT in various applications is the lack of reliable ways to quantify trustworthiness of IoT data~\cite{voas2018internet}. Data trust can be assigned to data as a qualitative attribute to help determine its reliability and integrity. Untrustworthy sensor data does not represent the reality accurately and may originate from a variety of unfaithful sources~\cite{Jayasinghe2017}. Untrustworthy data may be created intentionally (e.g., by malicious actors) or unintentionally (e.g., due to system failures or unexpected changes in circumstances).

In the context of IoT, data trust can be associated with a single value or a series of values captured by a sensor. Given a reliable trust score, data can be vetted before being used in any capacity. This can ensure data integrity and, consequently, improve the usefulness of the information inferred from the data~\cite{Hussain2020}.
In IoT systems, interactions at three layers can affect data trustworthiness. The first layer is the perception layer where IoT devices are used to sense and gather data. The second layer is the network layer, which facilitates communication with other IoT devices or servers. The third layer is the application layer that pertains to data processing and service delivery to the users~\cite{Mahmoud2015}. In this paper, we address data trust at the perception layer where sensor data is susceptible to noise, error, or malicious manipulation~\cite{DACHYAR2019e02264,Mahmoud2015}.

Previous research include several approaches to IoT data trust evaluation. For example, researches have used blockchain to asses the trustworthiness of sensor data based on data correlation, sensor node reputation, and data confidence in an immutable manner~\cite{Dedeoglu2019}. Some researchers have proposed to evaluate IoT data trust by analyzing suspicious network behavior~\cite{Mahmud2018, Reddy2017, Chinnaswamy2021} or by examining raw sensor data~\cite{Banerjee2018, Jayasinghe2019, Karmakar2020}.
However, most of these approaches evaluate the trustworthiness of sensor nodes~\cite{Raya2008,Momani2007} and not directly that of the sensor data itself. It is important to evaluate the trustworthiness of data as IoT sensors may produce both trustworthy and untrustworthy data at different occasions or circumstances.

Several existing works on evaluating IoT sensor data trust utilize supervised machine learning (ML) algorithms to classify the data in terms of trustworthiness in a binary fashion~\cite{Banerjee2018, Karmakar2020, Jayasinghe2019}, i.e., trustworthy versus untrustworthy.
To the best of our knowledge, there is no real-world IoT dataset that contains both trustworthy and untrustworthy data with ground-truth trust labels. Therefore, it is common to use clustering algorithms to artificially generate trust labels from unlabeled data.
In addition, since raw time-series sensor data is usually high-dimensional, it is typical to represent the data using lower-dimensional features that are extracted from the raw data and facilitate the classification of its trustworthiness. The features most commonly used in the existing related works are based on the Dempster-Shafer theory (DST)~\cite{Reddy2017, Chinnaswamy2021, Banerjee2018, Karmakar2020}.
Thus far, the most common ML-based approach to IoT data trust evaluation has been to use a classifier, such as support vector machine (SVM), that is trained on data labeled via $k$-means clustering~\cite{Jayasinghe2019, Banerjee2018, Karmakar2020}. However, due to the lack of real-world datasets with ground-truth trust labels, the true efficacy of this approach is not clear. One of our primary objectives is to shed some light on this approach.

Our main contributions in this paper are as in the following.
\begin{itemize}
    \item We develop a new method to synthesize untrustworthy data and augment real-world datasets containing time-series IoT sensor data. Our goal is to create datasets whose datapoints are meaningfully and accurately labeled as either trustworthy or untrustworthy.
    \item We propose to extract a new set of features from time-series IoT sensor data that capture the spatiotemporal correlations in the data.
    \item We show that the proposed correlation-based features can lead to better accuracy in classifying data trustworthiness compared to the DST-based features commonly used in the literature.
    \item We run extensive experiments using a real-world IoT sensor dataset with ground-truth trust labels, which we augment using synthesized untrustworthy data. The results show that the popular approach of SVM classification in tandem with labeling through $k$-means clustering is not as effective as presumed in previous works. Therefore, we challenge the prevailing assumption that clustering can yield meaningful IoT data trust labels.
    \item We reveal the promising potential of semi-supervised ML in learning accurate IoT data trust classification models without relying on extensive labeled data.
\end{itemize}

We organize the remainder of the paper as follows. We summarize the most relevant existing works in section~\ref{sec:relatedwork}. We describe our new method for synthesizing untrustworthy data in section~\ref{sec:dataset} and the calculation of our new correlation-based features in section~\ref{sec:featuresselection}. We examine various ML-based approaches to IoT data trust evaluation in section~\ref{sec:Evaluation}. Finally, we discuss our main findings in section~\ref{sec:Discussion} and conclude the paper in section~\ref{sec:Conlusion}.

\section{Related Work} \label{sec:relatedwork}

We present the existing works on IoT data trust evaluation within two categories, the conventional approaches that are mainly based on heuristics and the data-driven approaches that are based on ML.

\subsection{Conventional approaches}

A variety of techniques have been proposed for evaluating IoT data trust. Approaches utilizing weighted voting, Bayesian inference, and DST have demonstrated good performance. However, they often do not directly evaluate the trustworthiness of the data, but that of the sensor from which the data emanates~\cite{Raya2008}. This is limiting as IoT sensors may not always produce data that is solely trustworthy or untrustworthy. The trustworthiness of IoT data can vary depending on various factors such as sensor reliability, environmental conditions, data transmission integrity, and potential cybersecurity threats.

The Bayesian and Gaussian reputation systems proposed in~\cite{Kurniawan2015, Singh2021, Lin2016}, which have been verified mostly using simulated network packet data, can provide some insights into data trust. However, reputation-based systems in the context of IoT naturally focus on evaluating the trustworthiness of IoT sensors rather than that of the data captured by the sensors~\cite{Momani2007}.

\subsection{ML-based approaches}

The works of~\cite{Jayasinghe2017, Jayasinghe2019} are among the first ones that use ML algorithms to assess trustworthiness of IoT data via binary classification. They consider extracting several features from raw sensor data that contain information related to the trustworthiness of data. The most commonly used features are the DST-based belief and plausibility values~\cite{Banerjee2018, Karmakar2020}. To calculate these features, time-series IoT sensor data is generally replaced by the probability mass function (PMF) of its values. However, the PMF of time-series sensor data may not fully capture all the information, especially regarding the sequential nature of the data. Time-series data typically exhibits temporal dependencies and patterns such as trends, seasonality, autocorrelation, or other temporal relationships. PMF alone does not consider the ordering or temporal dependencies between the datapoints.

Most existing ML-based approaches to evaluating trust in IoT data have been validated through experiments that can be considered contrived. One commonly used experimental setup involves learning an ML model, typically linear SVM, to classify the trustworthiness of the data after labeling it artificially using an unsupervised cluster analysis method, with the $k$-means clustering algorithm being the prevalent choice~\cite{Jayasinghe2019, Banerjee2018, Karmakar2020}. Such setups are adopted mainly due to the lack of ground-truth labels and the presumption that the trustworthy and untrustworthy data form distinct clusters in the feature space, which can be discovered via simple cluster analysis methods such as $k$-means clustering. The $k$-means clustering algorithm assumes multivariate Gaussian distributions with equal covariances for the clusters corresponding to the trustworthy and untrustworthy data. There is no evidence to support the validity of this assumption nor to substantiate the underlying presumption that clustering can provide reliable trust labels.

\section{Synthesis of Trust-Labeled Data} \label{sec:dataset}

\begin{figure*}
    \centering
    \includegraphics[scale=.45]{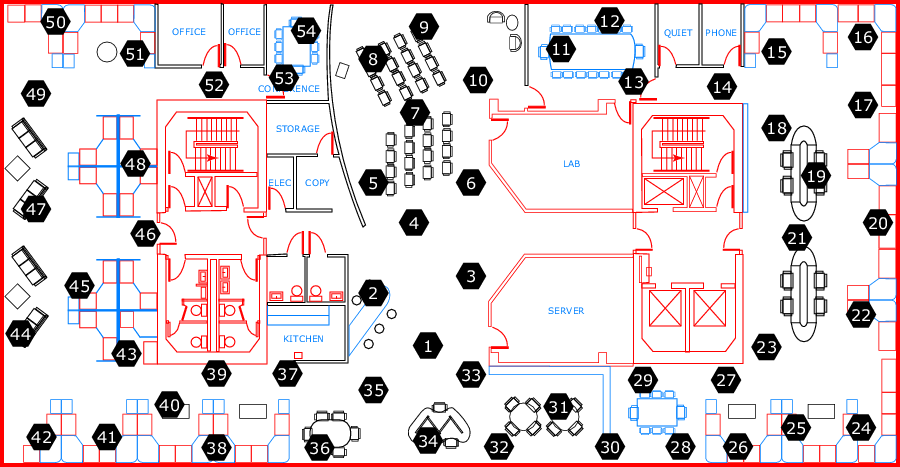}
    \caption{Intel Lab layout with the sensor locations.}
    \label{fig:sensor_locations}
\end{figure*}

In practice, acquiring large amounts of data with trust labels, specifically, data that is explicitly labeled as trustworthy or untrustworthy, is challenging. Collecting untrustworthy data that accurately reflects the types of data encountered in the real world can be particularly challenging. In many cases, data collection efforts focus on obtaining trustworthy and high-quality data for analysis and decision-making. As a result, untrustworthy data, which may include data with errors, noise, biases, or malicious manipulations, is less readily available or intentionally avoided.

The scarcity of trust-labeled data severely limits the utility of supervised ML algorithms for evaluating trust in IoT data. This has prompted many researchers to resort to using unsupervised clustering algorithms and labeling the trustworthiness of IoT data based on the inferred cluster assignments. However, in the absence of ground-truth labels, there is no assurance of the accuracy of the labels produced via clustering nor is there any reliable way to verify their veracity. Here, we propose a novel method to synthesize realistic untrustworthy IoT sensor data using real-world data collected through IoT sensor networks. The synthesized data can reasonably be considered untrustworthy as it does not originate from any real sensor measurement and is designed to deviate from the known trustworthy data in meaningful ways that can be practically expected. In other words, our synthetic untrustworthy data intentionally embodies deviations or errors, which can represent those that may occur in real world. As a result, it does not accurately reflect the complexities and nuances of real data and hence is untrustworthy.

To generate a trust-labeled dataset, we utilize the Intel Lab dataset that contains temperature readings from 54 sensors distributed around the Intel Berkeley Research Lab~\cite{bodik2004intel} (see Fig.~\ref{fig:sensor_locations}). We label the data of the original dataset as trustworthy except for conspicuous outliers, which have values three or more standard deviations away from the respective means. We propose a new algorithm, called random walk infilling (RWI), to synthesize untrustworthy data by altering sections of the original trustworthy time-series data. For comparison, we also consider an existing time-series data synthesis method proposed in~\cite{Banerjee2018}, which we refer to as \textit{Drift}.

\subsection{Random Walk Infilling}

The proposed RWI algorithm synthesizes untrustworthy data that shares similarities with the given trustworthy data but purposely deviates from it to an extent that renders it untrustworthy. We devise the algorithm such that it respects the time-series nature of the data despite utilizing random processes for data synthesis. Despite the intentional deviations, the synthesized untrustworthy data is designed to appear realistic, resembling the main characteristics and patterns present in the trustworthy data. Thus, RWI avoids synthesizing unrealistic data that, for example, may contain discontinuities, excessively abrupt changes, or dramatic drifts in the values over long sequences.

We implement the RWI algorithm through the following three main steps.
\begin{itemize}
    \item For the data of any sensor collected during given intervals, such as every day (24 hours), we extract the indexes (timestamps) of the first and last values and find a predetermined number of (e.g., ten) equally-spaced indexes between the first and last indexes.
    \item We replace the sensor data values within each segment, which lies between every pair of adjacent indexes, with synthetic values produced via a random walk process, where the step variance for the random walk is predetermined.
    \item We adjust the synthesized values within each segment to approximately align the start and end values of the segments with the preceding and succeeding segments. We achieve the alignment by pivoting the synthesized values around the first value of each segment. This way, we preserve the long- and mid-term trends present in the original data and avoid introducing abrupt changes or discontinuities in the synthesized data, maintaining its overall continuity and coherence with the original data.
\end{itemize}

\begin{figure}
    \centering
    \includegraphics[scale=.55]{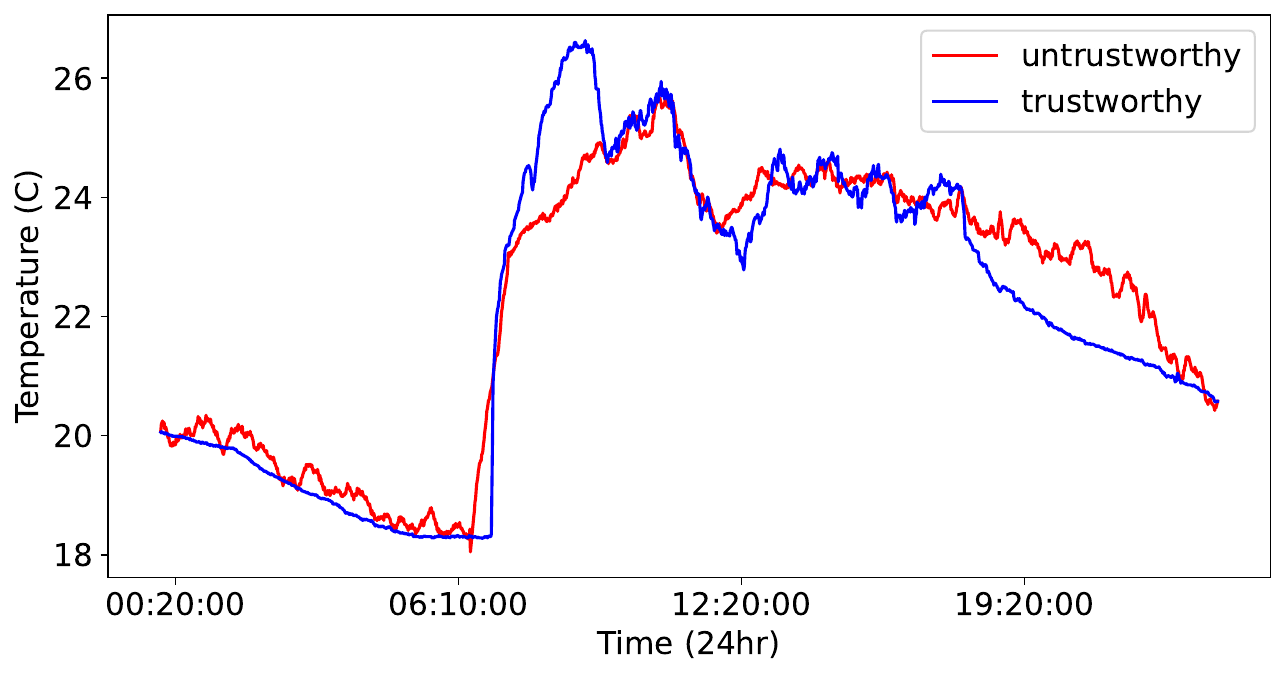}
    \caption{An example of actual trustworthy data instance (time-series data collected by a sensor in a 24-hour period) and a corresponding untrustworthy data instance synthesized via RWI.}
    \label{fig:randwalk_comparison}
\end{figure}

The resultant synthesized data has the same number of values as the original data. It roughly captures the non-short-term cyclic behaviors or trends in the original data without featuring any improbable discontinuity, sudden change, or undue trend shift. We summarize the main computational steps of the proposed RWI algorithm in Algorithm~\ref{alg1}. In Fig.~\ref{fig:randwalk_comparison}, we plot the values of an example 24-hour-long time-series data instance, labeled as trustworthy, and its corresponding synthetic counterpart created via RWI and labeled as untrustworthy.

Note that to better preserve the trends of the original data within the synthetic data, one can find the maxima and minima in the original data and use them as the segment boundaries. However, in our experience, choosing a sufficiently large number of equally-spaced points is adequate to maintain the trends to a satisfactory level. 

\begin{algorithm}
\caption{Random walk infilling.}
\textbf{input} dataset containing instances corresponding to the times-series data of each sensor and given interval (e.g., 24 hours); random-walk step variance $\sigma^2_g$; number of middle points $M$\\ 
\textbf{output} synthesized dataset $\mathcal{A}$
\label{alg1}
\begin{algorithmic}
\State initialize $\mathcal{A} \gets \emptyset$
\For {every instance in the input dataset}
    \Statex\quad\ find the indexes of the first and last values and $M$ mid ones:
    \State $\mathcal{P} =\{p_0, ..., p_{M+1}\}$
    \For {every segment $i=0,\dots,M$}
        \Statex\quad\quad\quad find the segment values indexed between $p_i$ and $p_{i+1}$:
        \State $\mathcal{S} \gets \{s_{p_i}, ..., s_{p_{i+1}}\}$
        \Statex\quad\quad\quad calculate the segment slope (inclination):
        \State $b=\frac{\sum_{j=p_i+1}^{p_{i+1}}{(j-p_i)(s_{j}-s_{p_i})}}{\sum_{j=p_i+1}^{p_{i+1}}{(j-p_i)^2}}$
        \Statex\quad\quad\quad replace the values with synthetic ones via random walk:
        \For {$j=p_i+1,...,p_{i+1}$}
            \State $s_j \gets s_{j-1}+g_j$, $g_j \sim \mathcal{N}(0,\,\sigma_g^{2})$ 
        \EndFor
        \Statex\quad\quad\quad calculate the slope difference:
        \State $b \gets b-\frac{\sum_{j=p_i+1}^{p_{i+1}}{(j-p_i)(s_{j}-s_{p_i})}}{\sum_{j=p_i+1}^{p_{i+1}}{(j-p_i)^2}}$
        \Statex\quad\quad\quad pivot the synthesized values to restore the original slope:
        \For {$j=p_i+1,...,p_{i+1}$}
           \State $s_j \gets s_j + b(j-p_i)$ 
         \EndFor
        \Statex\quad\quad\quad update the synthesized dataset:
        \State $\mathcal{A} \gets \mathcal{A} \cup \mathcal{S}$
    \EndFor
\EndFor
\end{algorithmic}
\end{algorithm}


\subsection{Drift}

\begin{figure}
    \centering
    \includegraphics[scale=.55]{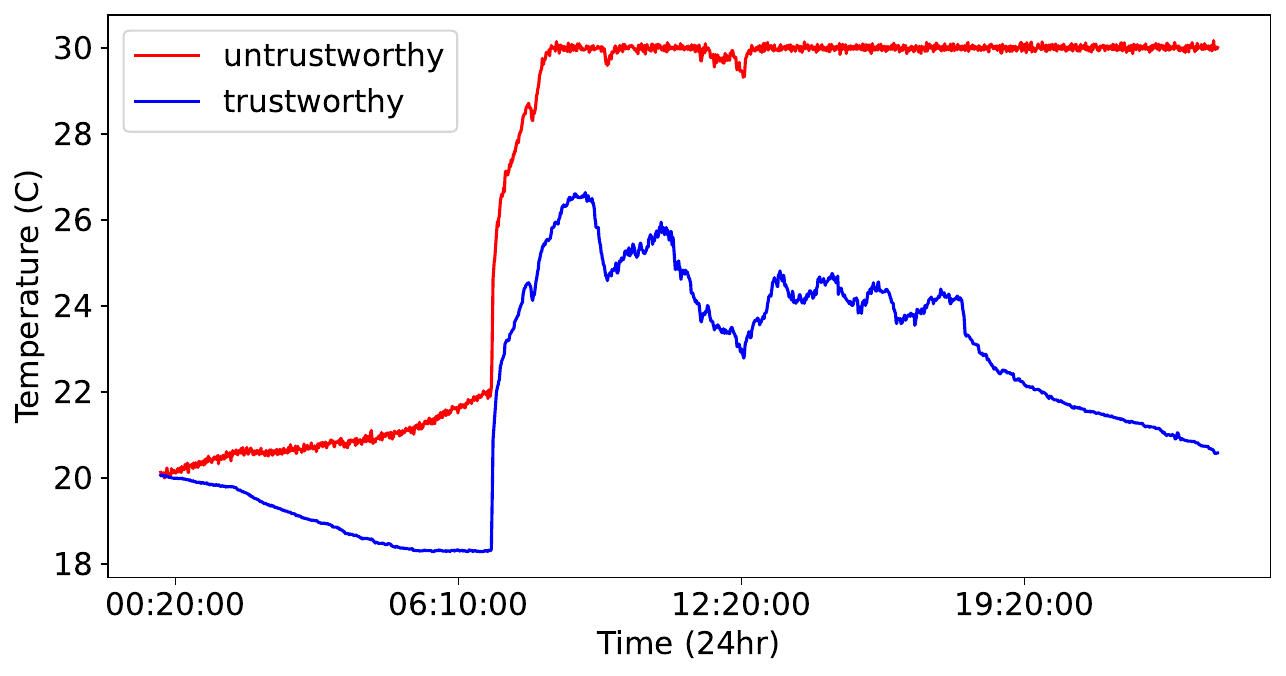}
    \caption{An example of actual trustworthy data instance (time-series data collected by a sensor in a 24-hour period) and a corresponding untrustworthy data instance synthesized via Drift.}
    \label{fig:driftgen_comparison}
\end{figure}

The Drift data synthesis method employed in previous research~\cite{Banerjee2018, Karmakar2020} generates a specific type of untrustworthy data that can typically be caused by a faulty sensor or a malicious actor. It generates synthetic data by cumulatively adding a so-called drift factor to the original time-series data until an upper limit is reached. The drift factor is composed of two terms, a constant term and an additive Gaussian noise term. In Fig.~\ref{fig:driftgen_comparison}, we plot the values of an example synthetic data instance (time-series data of a sensor during a 24-hour period) generated via Drift together with the original data instance. Comparing the synthetic data created via RWI and Drift in Figs.~\ref{fig:randwalk_comparison} and~\ref{fig:driftgen_comparison} shows that the Drift method may result in synthetic data that is more distinctly different from the original data compared with the RWI method. This salient difference indicates that the synthetic data generated by Drift may be less realistic or excessively artificial while being rightfully untrustworthy. In essence, due to its simplistic and monotonic manipulation of the original data, Drift may introduce deviations that are inconsistent with the underlying patterns or behaviors in the original data. This can result in synthetic data instances that do not accurately represent the wide range of untrustworthy data that only subtly differ from the trustworthy data.

\section{Features} \label{sec:featuresselection}

Extracting informative features from high-dimensional raw sensor data, that can effectively represent the data in a lower-dimensional subspace, is critical for creating accurate and efficient ML models. Here, we consider calculating two sets of features for time-series data. The features of the first set capture the spatiotemporal correlations in the data. The second set of features are based on the DST theory as proposed in~\cite{Banerjee2018, Karmakar2020}. We calculate the features over two-hour time-series data windows. We also limit the neighborhood of each sensor to its seven closest neighbors.

We determine the proximity of neighboring sensors using two factors, the physical (Euclidean) distance and the historical correlation. We initially consider the fifteen physically closest sensors to be the neighbors of any given sensor. We then reduce the number of the neighbors to seven by selecting the ones with the highest historical Pearson correlation score. We use the selected seven neighboring sensors to evaluate the spatiotemporal correlation features as well as the belief and plausibility features pertaining to DST.

\subsection{Correlation Features}\label{ssec:corfeat}

We hypothesize that trustworthy and untrustworthy data can be accurately distinguished using features that capture data auto-correlation and cross-correlation. For the time-series data of any given time window, the auto-correlation features that we calculate relate to the representation of the data in the frequency domain and our cross-correlation features are the Pearson's correlation coefficients of the data and those of the neighboring sensors.

\subsubsection{Frequency Domain Representation}

Time-series data can be represented in the frequency domain using the coefficients of its discrete cosine transform (DCT).
Let $x_i$ be the $i$th value of a given time window with $N$ values. We calculate the DCT coefficients as~\cite{DCT}
\begin{equation}
    a_k =\sum_{i=0}^{N-1} x_i \cos\left[\frac{\pi}{N} \left(i + \frac{1}{2}\right)k\right] \text{ for }\ k = 0,\dots,M-1
\end{equation}
where $M$ is the number of DCT coefficients, which is constant for the data of all time windows. We then split the coefficients into ten equal frequency bands and average them within each band. The resultant ten average values are our auto-correlation features that can compactly represent the data in the frequency domain. Note that these coefficients do not strictly represent the data auto-correlation that is defined as the correlation of a time-series data with a delayed copy of itself as a function of the delay. Rather, they represent the correlation of the data with itself from a frequency-domain point of view.

\subsubsection{Correlation with Adjacent Sensors}

We calculate the cross-correlation features, as the Pearson's correlation coefficients between the sensor data and the data of the neighboring sensors, via
\begin{equation}
    b_n = \frac{\sum_i(x_i-\bar{x})(y_{n,i}-\bar{y_n})}{\sqrt{\sum_i(x_i-\bar{x})^2\sum(y_{n,i}-\bar{y_n})^2}}
\ \ \text{for}\ n = 1,\dots, 7
\end{equation}
where $y_{n,i}$ is the $i$th data value of the $n$th neighboring sensor corresponding to the sensor data $x_i$ and $\bar{x}$ and $\bar{y}_n$ are the means of $\{x_i\}_{i=1}^N$ and $\{y_{n,i}\}_{i=1}^N$, respectively. Assuming that $x_i$ and $y_{n,i}$ arise from jointly wide-sense stationary stochastic processes, $b_n$ features can be viewed as the normalized cross-covariance values for lag zero.

We concatenate the above-mentioned ten auto-correlation and seven cross-correlation features to form a seventeen-dimensional feature vector corresponding to the data of each two-hour time window.

\subsection{DST Features}

Based on the assumption that the probabilistic nature of DST can help recognize the trustworthiness of time-series data, features extracted from the data through DST are commonly used for evaluating data trust in IoT applications. There are several DST-based features that can be calculated from IoT sensor data. In most recent works on evaluating IoT data trust, the DST-based features are calculated using a measure of distance between the belief and plausibility values of the data and those corresponding to the data of the neighboring sensors. Here, we calculate these features using the Canberra distance as in~\cite{Banerjee2018, Karmakar2020} while considering seven neighboring sensors.

\subsection{Feature Space Visualization} \label{sec:featurespacevisual}

\begin{figure}
    \centering
    \includegraphics[scale=.55]{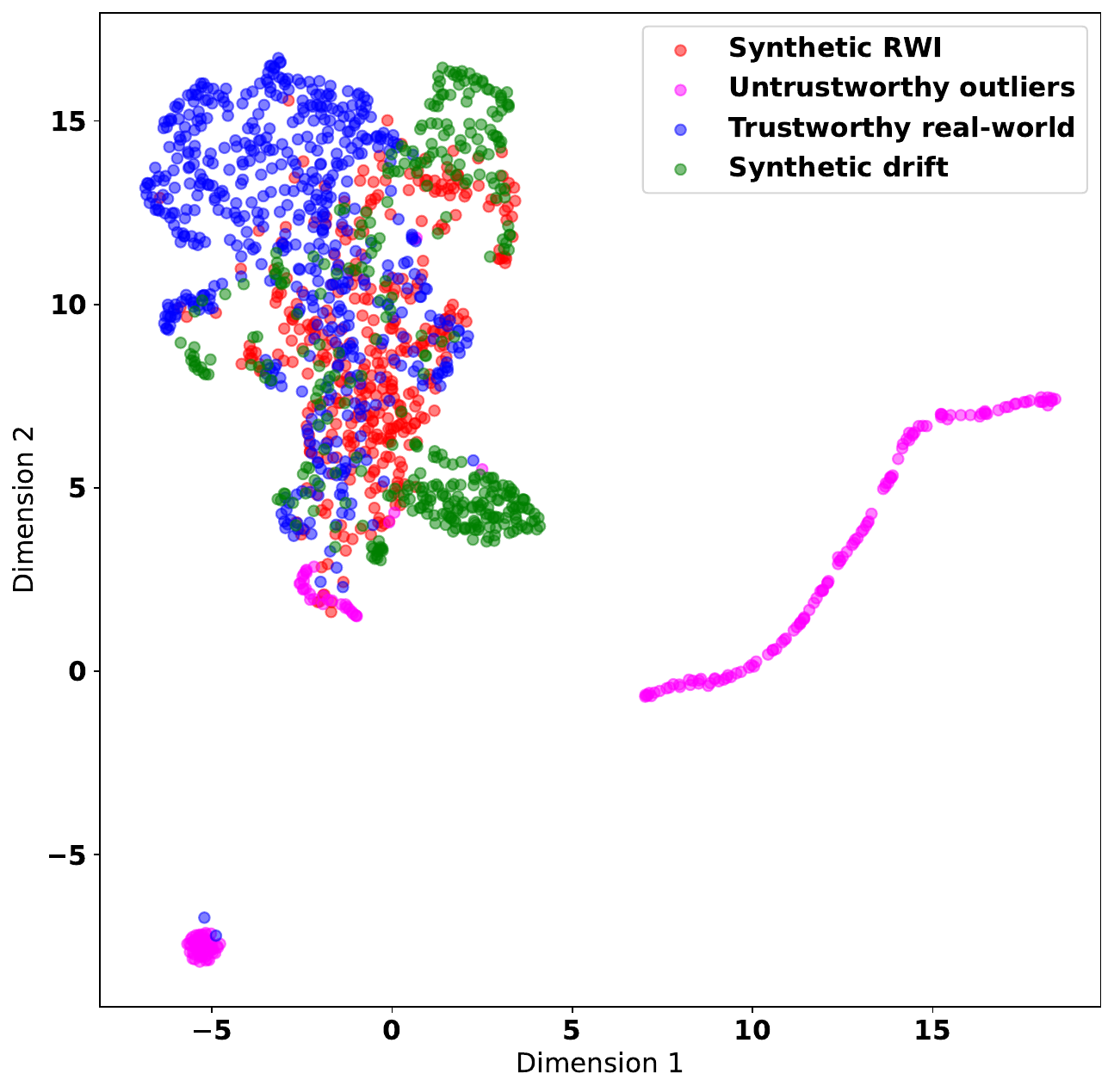}
    \caption{Visualization of the correlation feature space of the Intel Lab dataset, which is augmented by synthesized RWI and Drift untrustworthy data, using the UMAP algorithm.}
    \label{fig:Feature_vis_RWI}
\end{figure}

We visualize the correlation features (as described in section~\ref{ssec:corfeat}) that we extract from the Intel Lab dataset. We augment the dataset by adding untrustworthy data synthesized through both RWI and Drift methods. We utilize the uniform manifold approximation and projection (UMAP) algorithm~\cite{umap} for the visualization of the feature values, which originally reside in a seventeen-dimensional feature space, in a two-dimensional embedding subspace (plane). We show the results in~\ref{fig:Feature_vis_RWI}. The figure also includes the original data instances that are conspicuously erroneous or abnormal. We consider these instances to be untrustworthy and label them as untrustworthy outliers. 

Note that UMAP makes three assumptions about the data, namely, (i) the data is uniformly distributed on a Riemannian manifold; (ii) the metric, which quantifies the distance between data points, is locally constant; and (iii) the manifold representing the underlying structure of the data is locally connected. UMAP preserves the local structure of the data as it produces a data embedding where the relationships and clusters in the original data are maintained. This embedding aids in visually identifying clusters and understanding the organization of the data.

In~Fig.~\ref{fig:Feature_vis_RWI}, the untrustworthy data instances synthesized by the Drift method appears to be further apart from the trustworthy data instances in the correlation feature space compared with the untrustworthy data instances synthesized by the RWI method. This implies that the RWI synthetic data more closely resemble the original trustworthy data compared with the Drift synthetic data. Therefore, RWI synthetic data is potentially more realistic, but also more challenging to distinguish from the trustworthy data.
In addition, there appears to be regions in the correlation feature space where the trustworthy data does not mix with the synthesized untrustworthy data. Classifying the trustworthiness of data is likely to be easier in regions where the trustworthy data and untrustworthy data form separate clusters, as opposed to regions where the two types of data are mixed and do not exhibit distinct clusters.

\section{Evaluation} \label{sec:Evaluation}

In this section, we present some numerical simulation results to evaluate the effectiveness of the proposed RWI synthesis method and the proposed correlation features. In addition, we shed some light onto the challenges of developing meaningful IoT data trust evaluation models and the issues surrounding some existing approaches that attempt to circumvent the limitation of the lack of ground-truth-labeled untrustworthy data through unsupervised clustering. We describe the ML-based approaches that we consider for classifying trustworthiness of IoT data, the related algorithms, the utilized data, and a brief justification of the considered approaches/algorithms. We then summarize the evaluation results for all considered approaches in terms of data trust prediction accuracy whilst drawing attention to our main findings discussed in the succeeding sections.

\subsection{Considered ML-based approaches}

In the literature, both supervised and unsupervised ML algorithms have been used to classify the trustworthiness of IoT data or sensors/devices. Supervised ML algorithms are trained for classification by minimizing the error between the predictions and the ground truth. Semi-supervised ML algorithms can also be utilized for learning classification models from a mixture of labeled and unlabeled data. Unsupervised ML (clustering) algorithms are generally not designed for classification, but instead are useful for understanding existing patterns and relationships within the data. However, they can be used to classify unseen data based on their cluster assignment. After training a clustering algorithm using data with no class label, the data instances assigned to each learned cluster can be presumed to have the same class label. Unseen data can then be classified according to their proximity/distance to different clusters. This way, we can compare the classification accuracy of supervised and unsupervised ML algorithms when used to evaluate data trustworthiness.

The lack of publicly-available benchmark datasets makes the comparison of the existing approaches for IoT data trust evaluation difficult. Therefore, we base our comparisons on augmented versions of the Intel Lab dataset that contain untrustworthy data synthesized via the methods described in section~\ref{sec:dataset}. 

In our experiments, we consider the relevant approaches that are based on supervised, unsupervised, and semi-supervised ML. They include the fully-supervised approach based on learning a binary classification model from a dataset containing ground-truth data trust labels and the unsupervised approach of learning a classification model from a dataset whose data trust labels are obtained via clustering as proposed in~\cite{Jayasinghe2019, Karmakar2020}. We use two classification models, namely, linear SVM and multi-layer perceptron (MLP) with a single hidden layer. For cluster analysis, we use the $k$-means and Gaussian mixture model (GMM) clustering algorithms. In addition, we consider a semi-supervised approach based on the label propagation algorithm and evaluate its performance over partially-labeled datasets.

We compare the cross-validated performance of the considered ML-based approaches, in terms of the accuracy of data trustworthiness classification, over multiple datasets augmented by different methods and represented by different feature sets. We also evaluate the classification accuracy of each approach on datasets that differ from the one used for training. The purpose of this evaluation is to assess how well each approach generalizes to unseen data.

\subsection{Summary of results}

\begin{figure}
    \centering
    \includegraphics[scale=.55]{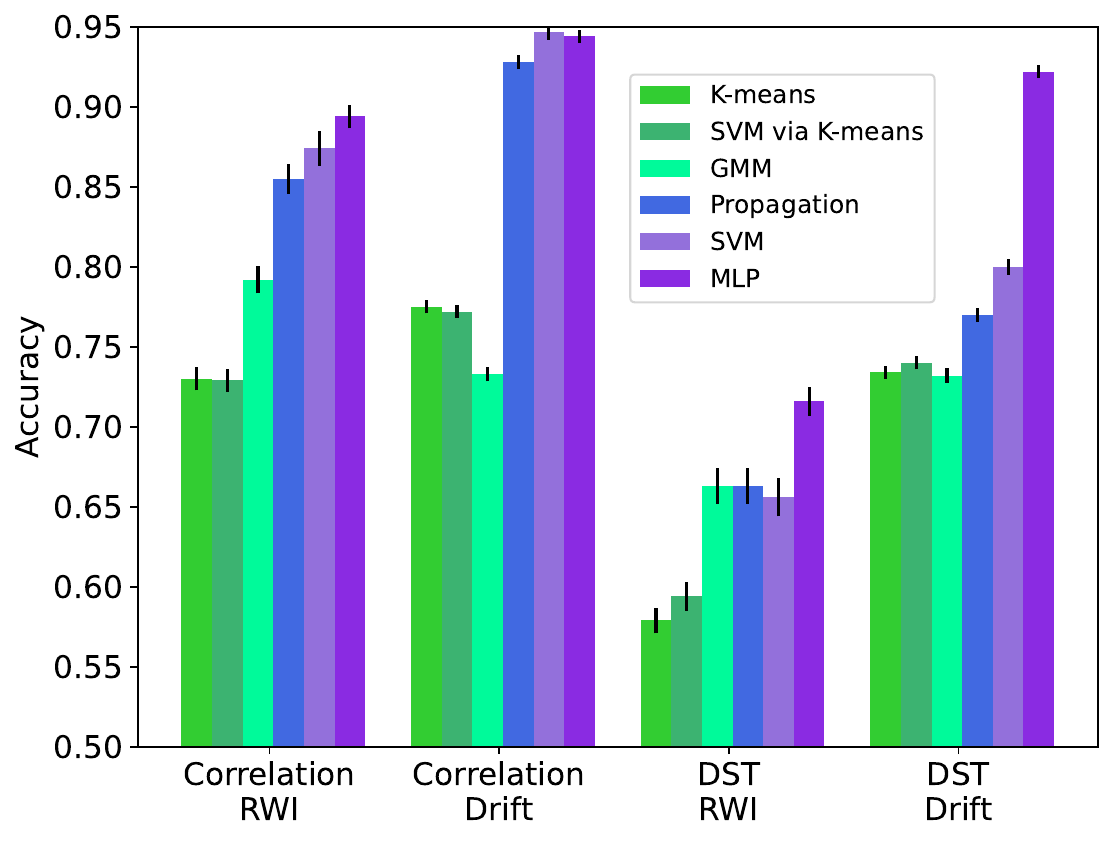}
    \caption{Data trust classification accuracy of the considered ML-based approaches trained and evaluated on both correlation and DST feature sets that are extracted from the Intel Lab dataset containing untrustworthy data synthesized using the RWI or Drift methods. The results are averaged over ten independent random realizations of data synthesis.}
    \label{fig:feature_perf}
\end{figure}

We evaluate the accuracy of the considered ML-based IoT data trust classification approaches via ten-fold cross validation and present the results in Fig.~\ref{fig:feature_perf} for both the correlation and DST features (section~\ref{sec:featuresselection}) extracted from the Intel Lab dataset that is augmented using both the RWI and Drift data synthesis methods (section~\ref{sec:dataset}). 
We calculate the accuracy as the total number of correct classifications over the total number of classification inferences for ten datasets containing independently generated synthetic untrustworthy data. In addition, we calculate the standard deviation of the accuracy over the ten independently-synthesized datasets for each case and present them as error bars in Fig.~\ref{fig:feature_perf}. 

\begin{figure}
    \centering
    \includegraphics[scale=.55]{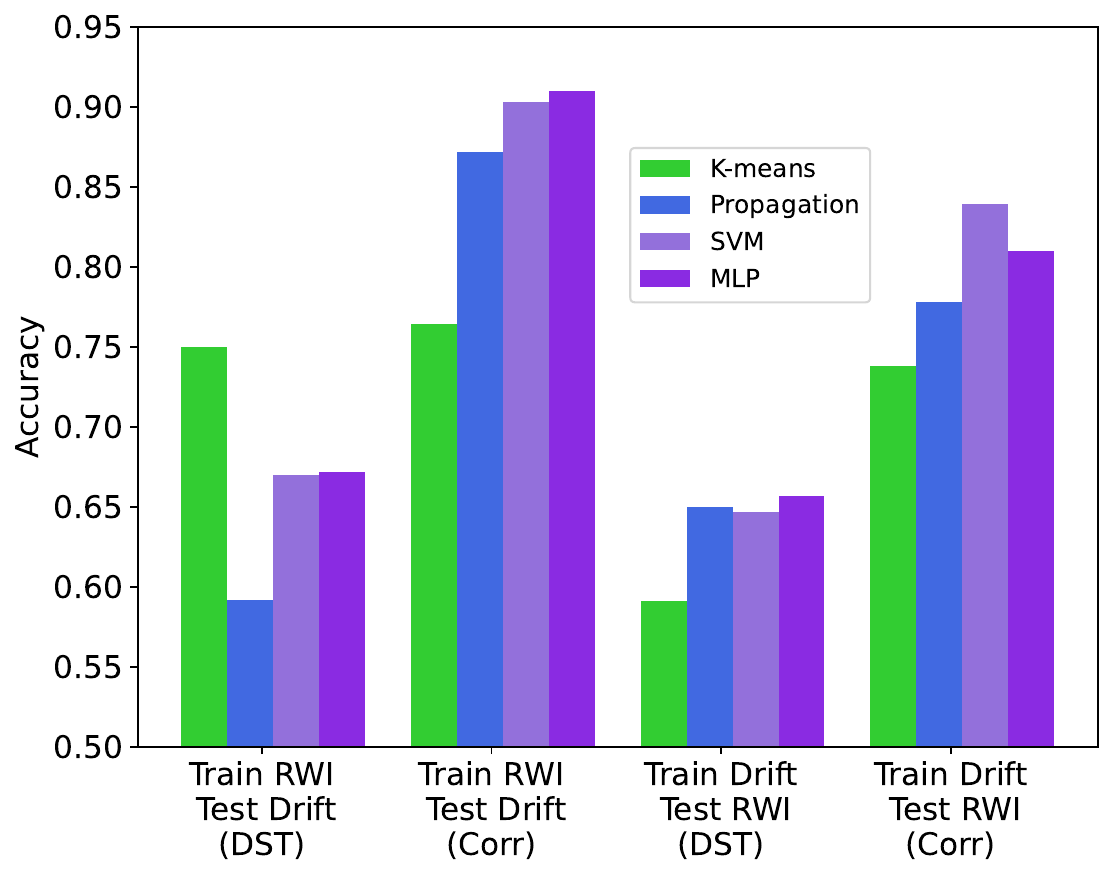}
    \caption{Data trust classification accuracy of the considered ML-based approaches when evaluated on unseen data, i.e., trained on a dataset containing untrustworthy data that is synthesized using either RWI or Drift method and tested on a dataset containing untrustworthy data synthesized using the other method.}
    \label{fig:dataset_perf}
\end{figure}

In Fig.~\ref{fig:dataset_perf}, we assess the ability of the trained models to generalize to data not seen during training. To this end, we provide the classification accuracy results for when the considered approaches are trained on a dataset augmented via either the RWI or Drift data synthesis method and tested on a dataset augmented via the other data synthesis method. We present the results for both the correlation and DST features.

\begin{figure}
    \centering
    \includegraphics[scale=.55]{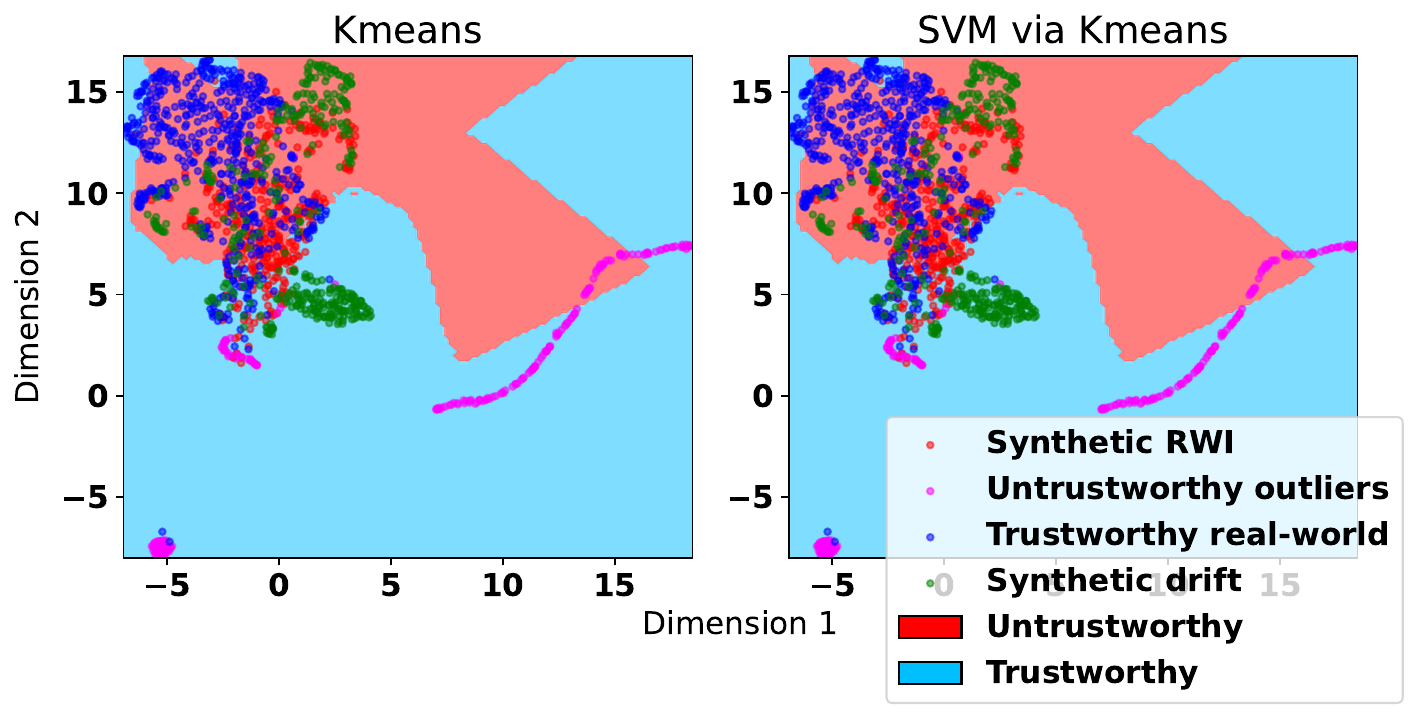}
    \caption{UMAP visualization of the decision boundaries learned via $k$-means clustering (left) and an SVM classifier trained on the data labeled via $k$-means clustering (right) when using a dataset augmented via RWI and the correlation features.}
    \label{fig:kmeans_svm_boundary}
\end{figure}

\begin{figure}
    \centering
    \includegraphics[scale=.55]{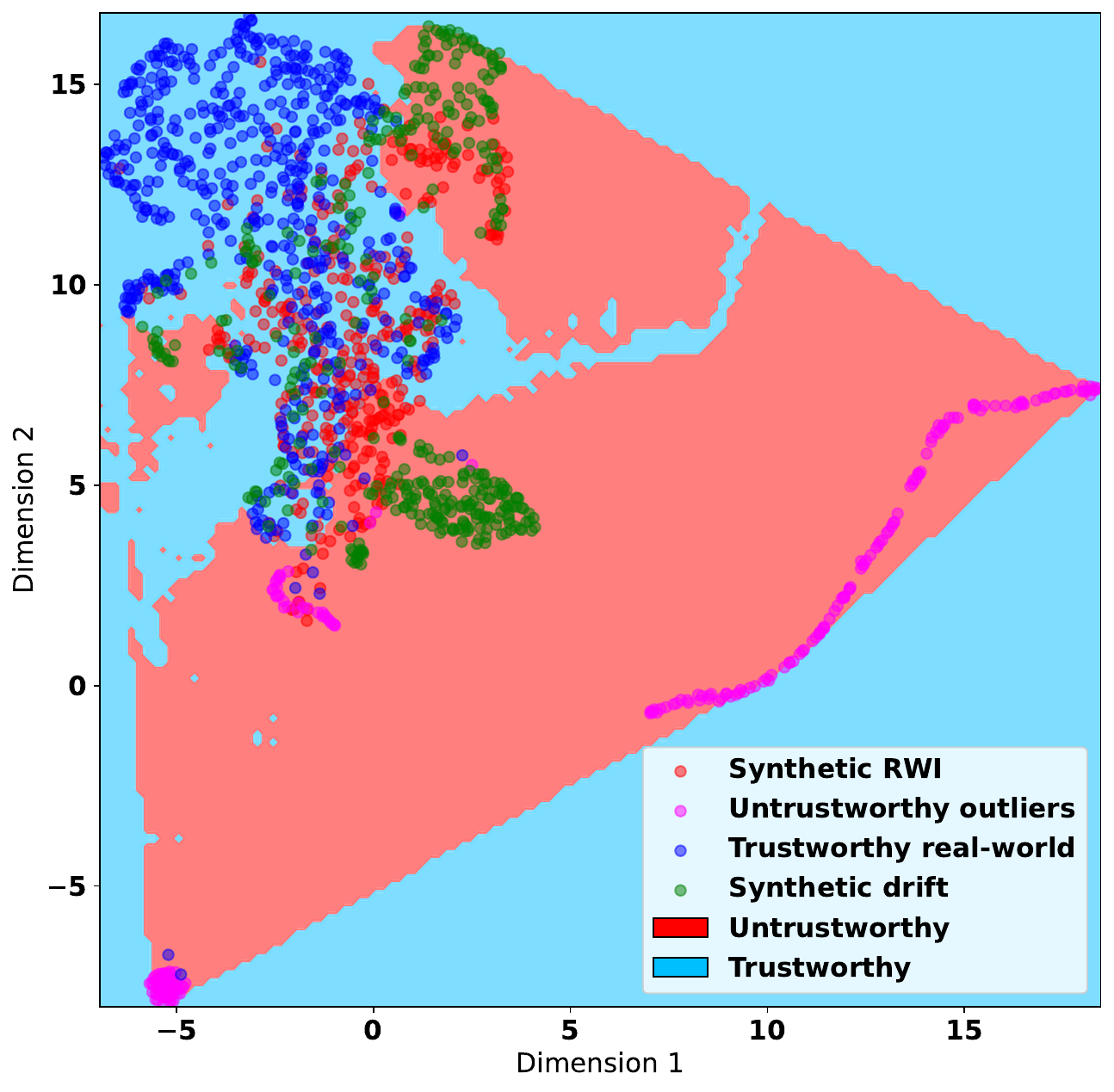}
    \caption{UMAP visualization of the trustworthy and untrustworthy decision regions in the correlation feature space determined by the MLP algorithm trained on a dataset containing untrustworthy data synthesized via the Drift method.}
    \label{fig:mlp_drift}
\end{figure}

\begin{figure}
    \centering
    \includegraphics[scale=.55]{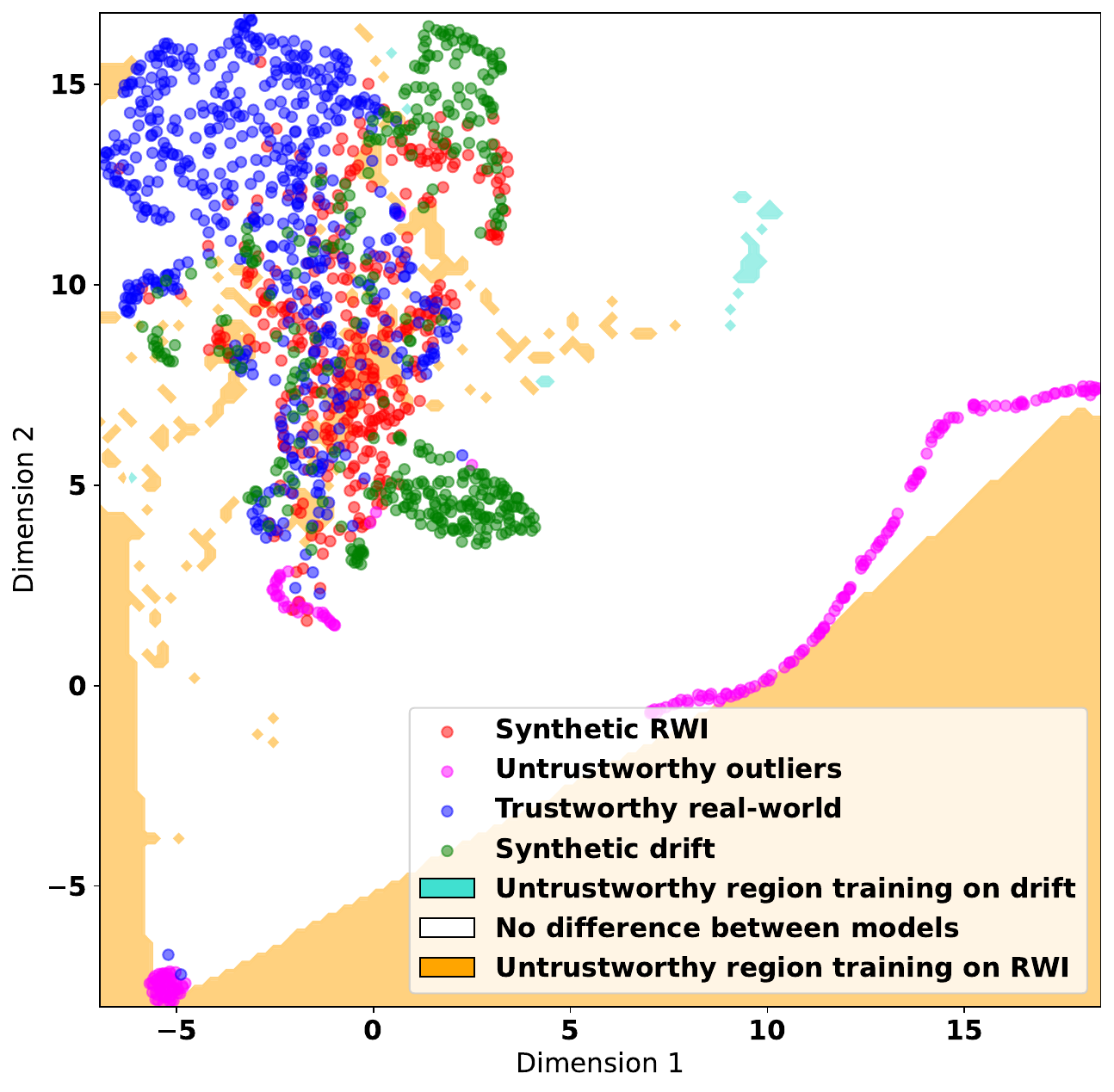}
    \caption{UMAP visualization of the unique trustworthy and untrustworthy decision regions in the correlation feature space determined by the MLP algorithm trained on datasets augmented by untrustworthy data generated via the RWI or Drift method.}
    \label{fig:mlp_drift_RWI_diff}
\end{figure}

As seen in Figs.~\ref{fig:feature_perf} and~\ref{fig:dataset_perf}, the correlation features yield higher accuracy compared to the DST features in all examined cases. Unsurprisingly, the supervised ML-based approaches (tagged as SVM and MLP) generally deliver the highest accuracy. The semi-supervised ML-based approach utilizing the label propagation algorithm (tagged as propagation) performs close to SVM and MLP, especially when using the correlation features and data synthesized using RWI. The unsupervised ML-based approaches (tagged as $k$-means, SVM via $k$-means, and GMM) have significantly lower accuracy compared with the supervised ML-based approaches. 
The standard deviation of the accuracy values due to randomized data synthesis is relatively low, with the largest value being 1.2\% for SVM using the correlation features and the RWI data synthesis method.

It is evident from Fig.~\ref{fig:dataset_perf} that the correlation features offer better generalizability compared to the DST features, in terms of the cross-dataset data trustworthiness classification accuracy. In addition, when using the correlation features, training data trust classification models on datasets augmented by RWI and evaluating them on datasets augmented by Drift results in higher accuracy for all considered ML-based approaches compared to the alternative, that is, training on datasets augmented by Drift data and testing on datasets augmented by RWI. Furthermore, training on datasets containing RWI synthetic data while using the correlation features results in the highest cross-dataset accuracy for all considered ML-based approaches.

\section{Discussion} \label{sec:Discussion}

In this section, we first examine the performance evaluation results regarding the ML-based approaches and features proposed in the previous related works. We particularly discuss the shortcomings of resorting to unsupervised cluster analysis for creating data trust labels, which are subsequently used to learn supervised classification models. We then discuss and contextualize the performance evaluation results specifically related to the usage of synthetic data generated via the RWI or Drift method. Finally, we compare the performance of all considered ML-based approaches and highlight the advantages of each approach.

\subsection{Previous Work}

A common existing approach to evaluate IoT data trust is to train supervised ML classification models, such as SVM, using datasets that are labeled via $k$-means clustering. That is, after clustering the unlabeled data, each data instance is given a label according to its assigned cluster. To shed some light onto the efficacy of this approach, we present the accuracy of prediction using the $k$-means cluster assignments as well as prediction using SVM trained on labels obtained from $k$-means clustering in Fig.~\ref{fig:feature_perf}. The results show that $k$-means and SVM via $k$-means deliver near identical accuracy with the difference in accuracy being not greater than 2\% for all considered feature sets and datasets.

Training an SVM classifier on data annotated with labels induced by $k$-means clustering can logically be less effective compared to training an SVM classifier on data with ground-truth labels. According to the results in Fig.~\ref{fig:feature_perf}, an SVM classifier trained on the labels inferred from $k$-means clustering yields similar accuracy to a classifier that uses the cluster boundaries determined by $k$-means clustering to infer the class label. This is because, in the absence of ground-truth data trust labels, an SVM or any other classification model fit to the clustering-induced labels is essentially redundant as it only approximates the cluster boundaries. To demonstrate this, in Fig.~\ref{fig:kmeans_svm_boundary}, we provide a UMAP visualization of the decision boundaries produced by $k$-means clustering and SVM classifier trained on labels predicted by $k$-means clustering when using the correlation features extracted from a dataset augmented via RWI. The figure also includes the representation of the utilized datasets as in Fig.~\ref{fig:Feature_vis_RWI}. We observe from Fig.~\ref{fig:kmeans_svm_boundary} that the decision boundaries yielded by the two approaches are very similar, which corroborates the futility of fitting any classifier, particularly linear SVM, over the data labeled via clustering.

In Fig.~\ref{fig:feature_perf}, we observe that the DST features result in less accurate models compared to the proposed correlation features, which can capture spatiotemporal patterns and correlations in the data. The DST features also yield lower generalization ability as seen in Fig.~\ref{fig:dataset_perf}. This is likely because the DST features do not take into account the temporal correlation within the data. Therefore, they cannot faithfully represent the temporal patterns, trends, or behaviors that may be indicative of the trustworthiness of the data. This is more prominent when the synthetic untrustworthy data is generated via the RWI method as seen in Fig.~\ref{fig:feature_perf}.

The DST features lead to improved performance when the untrustworthy data is synthesized via the Drift method. However, they are still outperformed by the correlation features on the same data. When the training and evaluation are performed on datasets augmented with untrustworthy data synthesized using the same method, the correlation features deliver higher accuracy compared with the DST features in all considered cases. Additionally, when using correlation features, models trained on datasets containing synthetic RWI data can classify unseen data synthesized via Drift with little loss of performance with respect to models trained on datasets containing synthetic Drift data. On the other hand, models trained on datasets augmented via the Drift method do not generalize well to datasets containing synthetic data generated by the RWI method, regardless of the feature set utilized. This attest to the superiority of RWI over Drift in terms of utility or effectiveness.

\subsection{Synthesis Methods}

Labeled untrustworthy IoT data can be challenging to obtain for several reasons. First, building labeled datasets for IoT data, especially in the context of data trustworthiness, can be resource-intensive and time-consuming. Collecting and annotating a significant amount of real-world IoT data with trustworthiness labels is often a complex and costly process. Second, determining the trustworthiness of IoT data can be subjective and context-dependent. Factors such as data sources, data quality, sensor reliability, communication issues, and environmental conditions can all influence the trustworthiness assessment. The lack of a universally agreed-upon definition or framework for labeling untrustworthy data adds to the complexity of acquiring labeled datasets. Third, IoT data often contains sensitive information related to individuals, organizations, or critical infrastructure. Labeling untrustworthy data may involve revealing vulnerabilities, security flaws, or potentially harmful patterns. This can raise privacy and security concerns, making it difficult to openly share or label such data for research or development purposes.

Synthesizing untrustworthy IoT sensor data that mimics real data but differs from it in subtle ways is a logical approach to obtaining labeled untrustworthy data. The ability to discern small discrepancies in data that indicate untrustworthy behavior can also inform our trust in more overt and less subtle behavior. In other words, being able to identify small red flags or inconsistencies in data can help us develop a better understanding of what constitutes trustworthy behavior. By extension, this knowledge can also aid us in assessing plainer (less subtle) traits that may be easier to recognize, but still require careful evaluation in order to establish trust.

Any data that does not directly emanate from real-world and accurate sources can be presumed untrustworthy. Drift in sensor readings is a well-known phenomenon that can compromise the trustworthiness of collected data. It can occur in the real world for various reasons, such as changes in environmental conditions, hardware or software malfunctions, or wear and tear of the sensor. It can manifest as perturbation/error/noise terms being added to the true values cumulatively as in the Drift method or in a more intricate temporally-localized manner as in the proposed RWI method.

Recalling our visualization of the datasets containing synthetic untrustworthy data in section~\ref{sec:featurespacevisual}, one can notice some patterns related to trustworthiness or untrustworthiness of data in Fig.~\ref{fig:Feature_vis_RWI}. We observe that many synthetic data instances created by Drift are considerably isolated from the trustworthy data, whereas the synthetic data created by RWI is somewhat intermixed with the trustworthy data. This means that the synthetic untrustworthy data generated by RWI is semantically more similar to the trustworthy data in the correlation feature space compared with the untrustworthy data synthesized by Drift. 

By construct, the proposed RWI synthesis method can be viewed as a generalization of the Drift method as it includes Drift as a spacial case where the data, using which the untrustworthy data is synthesized, is processed in a single segment.
This may explain why models learned using RWI synthetic data generalize well to Drift synthetic data while training on Drift synthetic data and evaluating on RWI synthetic data does not result in good accuracy as presented in Fig.~\ref{fig:dataset_perf}.
This observation also implicitly confirms that RWI synthetic data is untrustworthy despite its apparent resemblance to trustworthy data.

To gain some insights into the information that can be obtained through the untrustworthy data synthesized via the Drift and RWI methods, in Figs.~\ref{fig:mlp_drift} and \ref{fig:mlp_drift_RWI_diff}, we use the UMAP algorithm to visualize the decision boundaries learned by the MLP algorithm that split the feature space into trustworthy and untrustworthy regions. Fig.~\ref{fig:mlp_drift} corresponds to when the training dataset contains synthetic untrustworthy data generated by Drift and Fig.~\ref{fig:mlp_drift_RWI_diff} shows the difference in the decision regions due to training with Drift or RWI synthetic data. In both figures, we use the correlation features. They demonstrate that training on RWI synthetic data results in the identification of a larger untrustworthy region. This suggests that RWI synthesizes more comprehensive untrustworthy data that may encompass the data synthesized by Drift.



\subsection{ML Models}

The considered supervised ML-based approaches offer the best performance in classifying data trust using both RWI and Drift methods for synthesizing untrustworthy data and both correlation and DST feature sets. The best performing algorithm is MLP with the accuracy of 91\% when trained on RWI synthetic data and evaluated on datasets containing either RWI or Drift synthetic data as seen in Figs.~\ref{fig:feature_perf} and \ref{fig:dataset_perf}.

The considered unsupervised ML-based approaches perform relatively poorly in comparison with the supervised ML-based approaches. The considered semi-supervised ML-based approach utilizing the label propagation algorithm performs similar to the supervised ML-based approaches. This is encouraging as semi-supervised learning is more practical in IoT applications given that real-world IoT data is usually partially or sparsely labeled.

\section{Conclusion} \label{sec:Conlusion}

Evaluating the trustworthiness of data in IoT using supervised ML is promising but also challenging to realize due to the difficulty of obtaining data annotated with trust labels. We showed that synthesizing untrustworthy data with mostly similar characteristics to the available trustworthy data is an effective way of generating untrustworthy data and creating datasets that can be used to learn IoT data trust classification models. We proposed a new time-series sensor data synthesis method, called RWI, to generate untrustworthy data that closely resembles trustworthy data but is subject to subtle random deviations. To enhance IoT data trust evaluation performance, we also proposed calculating a new set of correlation features that can compactly represent the information of both temporal and spatial correlations in the data and help distinguish between trustworthy and untrustworthy data effectively.

Given the labeled synthetic data, we showed that when ground-truth trust labels are not available, contrary to the prior supposition, extracting labels from the data through clustering does not necessarily provide accurate or meaningful labels either for direct data trust inference or fitting an extra data trust classification model.
We also demonstrated the utility of a semi-supervised ML algorithm, i.e., label propagation, in evaluating IoT data trust using datasets in which only a small fraction of data is labeled. The considered semi-supervised approach delivered high accuracy comparable to that of supervised ML-based approaches. In real world, a semi-supervised ML-based approach is likely more desirable as it requires significantly less labeled data compared with those based on supervised ML.




\bibliographystyle{unsrt}  
\bibliography{main}

\end{document}